\documentclass[numberedappendix,twocolappendix]{emulateapj}
\usepackage{apjfonts}
\usepackage{natbib}
\usepackage{graphics}
\usepackage{multirow}
\usepackage{amsbsy}
\usepackage{mathrsfs}
\usepackage[normalem]{ulem}
\usepackage{amsmath}
\usepackage{xcolor}

\def\bicep{{\sc Bicep}}

\def\biceptwo{{\sc Bicep2}}

\def\planck{{\it Planck}}

\def\keck{\textit{Keck}}
\def\keckarray{\textit{Keck Array}}

\def\bk{\bicep/\keck}

\def\deg{^\circ}
\def\emode{$E$-mode}
\def\bmode{$B$-mode}
\def\BB{$BB$}

\newcommand{\cl}{\mathcal{C}_\ell}

\def\lcdm{$\Lambda$CDM}

\def\ttp{\textit{T}$\rightarrow$\textit{P}}
\def\etb{\textit{E}$\rightarrow$\textit{B}}

\def\alm{$a_{\ell m}$}

\def\aap{Astr. \& Astroph.\ }

\def\apj{Astrophys.\ J.\ }
\def\apjl{Astrophys.\ J.\ Lett.\ }
\def\apjs{Astrophys.\ J.\ Suppl.\ }

\def\prd{Phys.\ Rev.\ D\ }

\begin{document}

\title{Deprojecting beam systematics for next-generation CMB B-mode searches}
\author{Christopher Sheehy}
\email{chris.sheehy@descarteslabs.com}
\affiliation{Descartes Labs, New York, NY \& Brookhaven National Lab, Upton, NY}

\begin{abstract}

Measurements of the cosmic microwave background polarization are vulnerable to
systematic contamination from beam imperfections. Because the unpolarized CMB
$T$ is orders of magnitude larger than the polarized $E$ and $B$ signals, even a
tiny difference in instrument response between two orthogonally polarized
measurements of the CMB will result in a large non-zero differential signal,
even if the CMB is unpolarized.  Two strategies to mitigate this
temperature-to-polarization leakage are the use of a rotating half-wave-plate
and the fitting and removal of leakage templates from the polarized signal. The
half-wave-plate approach will, in principle, work for arbitrary beam shapes, but
in practice introduces complicated additional optics that themselves can
introduce systematics. The template deprojection approach is simple and requires
no additional hardware, but so far has approximated beam shapes as
elliptical Gaussians. In this work, we generalize the deprojection technique to
clean leakage from mismatch of arbitrarily shaped beams. We find that our technique
will clean leakage from main beam mismatch to the level of
$r\approx1\times10^{-5}$ without appreciable filtering of the cosmological
signal.

\end{abstract}


\maketitle
\section{Introduction}

As ground based measurements of the cosmic microwave background (CMB)
polarization become more sensitive, instrumental systematics must be controlled with
increasing stringency. Systematics that ``leak'' the
unpolarized temperature anisotropies into a polarized signal
are a particular source of concern because of the large amplitude of the
temperature anisotropies relative to the polarized signal.

A polarization measurement can be made by observing the same point on the sky with
multiple polarized detectors with multiple polarization angles as projected onto the
sky. The simplest of such measurements is performed by so-called ``pair
differencing.'' The signals from two orthogonally polarized detectors are
differenced to cancel the unpolarized $T$ component, leaving only a measurement
of the intrinsic polarization. If, however, the response of the two detectors
to the unpolarized component is at all mismatched, the bright temperature signal
is not fully canceled and the result is additive systematic contamination. The
mismatch may be the result 
of uncalibrated gain differences or optical beam mismatch. A general term for
this systematic is beam mismatch. Because a gain mismatch is
equivalent to a beam normalization mismatch, the techniques for dealing with
optical beam mismatch apply equally well to gain mismatch.

A method that has been successfully adopted to deal with \ttp\ leakage from
beam mismatch is given in \cite{BKIII} (hereafter BKIII). This
technique involves forming templates of \ttp\ leakage derived from the
\planck\ per-frequency $T$ maps (and their spatial derivatives) and fitting them
to the data. Though the optical beam is expected to remain constant in time, the
fitting has been done on short timescales to allow for time dependent gain
variation. The templates are a linear combination of the $T$ map and its first,
second, and cross derivatives. There are six such templates corresponding to the
modes of the difference of two elliptical Gaussians: differential gain (1),
width (1), centroid (2), and ellipticity (2). The small number of
templates relative to the number of degree-scale modes in the subset of
data over
which the fit is performed prevents excessive filtering of the cosmological
signal while completely filtering the \ttp\ leakage from the elliptical
Gaussian component of the mismatched beams. The filtering of the cosmological
signal over that produced by polynomial filtering and ground-fixed template
removal is small and, like those, manifests as a multipole dependent
transfer function. This transfer function is independent of the input signal and is empirically
determined by Monte Carlo simulations. The post-deprojection systematic
residuals are set by two factors: noise in the template map and the portion of leakage
arising from beam components not modeled by elliptical Gaussians.

As an additional check, BICEP/Keck makes in-situ beam maps of individual
detectors using a calibration source located on the ground and in the far-field
of the telescope. These beam maps are very high signal-to-noise \citep{BKIV} and
are convolved with the \planck\ $T$ map to predict the actual leakage in the
observed BICEP/Keck maps. Cross correlating the prediction with the observed map
estimates the amount of residual leakage after deprojection. The undeprojected
residual leakage estimate can also be included in the parameter estimation
procedure to produce unbiased estimates of the tensor-to-scalar ratio $r$ \citep{BKXI}. This,
however, relies on the fidelity of the beam maps and the forward simulation,
with increasingly stringent requirements as the sensitivity to $r$
increases. Furthermore, the beam maps themselves must be acquired through
extensive calibration campaigns during the short Austral summer, an increasingly
monumental task as the number of detectors climbs into the tens and hundreds of
thousands.

In this paper, we develop an extension to the deprojection technique to
remove leakage from arbitrarily shaped beams while leaving the true
cosmological signal mostly unfiltered. In Section~\ref{sec:sims}, we describe
our forward simulation of CMB observations and \ttp\ leakage. In
Section~\ref{sec:mapmaking}, we describe mapmaking and template generation
procedure. In Section~\ref{sec:results}, we report the results.

\section{Simulations}
\label{sec:sims}

We first generate sufficiently realistic simulations of time-ordered data (TOD)
at the per-detector level to test deprojection methods. Our simulations are a
simplification of those described in BKIII to simulate \ttp\ leakage in
\biceptwo\ and \keckarray\ data and to test subsequent mitigation methods. The
simplifications involved are primarily those of scan strategy, number of
detectors, and time stream filtering performed in addition to deprojection. We
make no simplifications that would result in simulated \ttp\ leakage that is
easier to deproject than in reality or that would affect our conclusions. The
main difference is that the beams we simulate are synthetic rather than the
actual measured beams of a real instrument.

The basic procedure is as follows: we produce multiple realizations of $T$, $Q$,
and $U$ sky maps that are then explicitly multiplied by a given detector's beam
at each point in the scan trajectory and summed. (Because the beams are not
circularly symmetric and because telescope boresight angle is not fixed,
producing a single beam-convolved input map is not possible.) Knowing the
detector's polarization angle, we then construct the signal as measured by the
detector at each point in time.

The simulated TODs are then passed to the mapmaking pipeline, which we describe
in Section~\ref{sec:mapmaking}. Like in BKIII, the \ttp\ leakage
component can be isolated by running the simulations with the input $Q$ and $U$
maps set to zero, in which case any non-zero polarized signal can only come from
\ttp\ leakage. In this section we detail these steps.

\subsection{Input Maps}\label{sec:inputmaps}

We construct 10 noiseless realizations of 143~GHz sky maps using the Python
\texttt{Healpy} routine \texttt{synfast}. The input CMB power spectra are
generated with the CAMB software\footnote{http://camb.info/} using the best fit
\lcdm\ model from \cite{P2013XVI}. (Using the more recent 2015 cosmological
parameters from \planck\ makes negligible difference.)  The lensing
$B$-mode~\citep{zaldarriaga98} is included by setting the input $BB$ power
spectrum to its expected value and, as such, does not contain off-diagonal
power. This is unimportant for the current study.  We also include \BB\ power
from galactic dust by adding to the \lcdm\ $\cl$'s the dust $\cl$'s appropriate
for $f_{sky} = 0.1$ at 143~GHz, extrapolated from the values provided in
in~\cite{PIPXXX} as described in \cite{Sheehy18}. This corresponds to $A_d =
6.125~\mu$K, the amplitude of the dust $\cl$ at $\ell=80$ and $f=$353~GHz, and
scaled to 143~GHz and to other $\ell$ using the power-law spectral indices
reported in \cite{PIPXXX}. Our Gaussian dust simulations are consistent with the
galactic dust properties in the \bk\ field~\citep{BKX}.  The Gaussian dust
simulations are thus only appropriate for relatively clean areas of sky.

Additionally, we produce ``$E$-only'' inputs maps by converting the maps to
\alm's, setting the \bmode\ \alm's to zero, and converting back to maps. We use
these maps only to assess the \etb\ mixing of our power spectrum estimator
later.

We must also include the effect of \planck\ measurement noise on the
deprojection templates. We use the \planck\ FFP8 noise simulations obtained from
NERSC\footnote{http://crd.lbl.gov/departments/computational-science/c3/} for the
143~GHz full mission maps~\citep{P2015XII}. We convolve the noiseless
simulated $T$ maps by the nominal \planck\ Gaussian beam and add to them the
FFP8 noise simulations to produce simulations of the maps ``as observed'' by
\planck. These serve as our deprojection templates. This procedure is the same
as that described in BKIII.

\subsection{Beams}\label{sec:beams}

We generate a discretized, simulated beam for each detector. We choose the grid
by generating a regular $(x,y)$ Cartesian grid with fixed spacing and converting
these to polar coordinates, $(r,\phi)$. We then treat the polar coordinates as
spherical coordinates and use them to compute the projected
R.A.\ and Dec.\ of each pixel in the discretized beam image given the beam
centroid's R.A.\ and Dec.\ and azimuthal orientation.  This procedure results in
the pixels of the gridded beam being slightly non-equal area, but this changes
only the interpretation of the gridded beam and does not affect the simulation
fidelity or deprojection efficiency.

We model individual detector beams as the sum of a circular Gaussian
and Gaussian-windowed Zernike polynomials. Different Zernike orders have explicit
azimuthal symmetry and therefore exhibit intuitive cancellation effects when
coadding over boresight angles as explained in BKIII. We first define a
Gaussian beam component common to all detectors,

\begin{equation}
B^g(r) = e^{-r^2/2\sigma^2},
\end{equation}

\noindent where $\sigma = 0.212\deg$ (corresponding to FWHM$=0.5\deg$). We then
construct a non-Gaussian component that varies from detector to detector. For the
$k$th detector this is

\begin{equation}
B_k^{ng}(r,\phi) = B^g(r) \sum_n \sum_{m=-n}^n a_{n,m} z_{n,m}(\rho,\phi),
\end{equation}

\noindent where $z$ is a Zernike polynomial of order $n$, $\rho \equiv r/1\deg$ is a
normalized radius coordinate, and $a_{n,m}$ are random coefficients drawn from a
Gaussian of mean 0 and width 1. The sum is over the orders
$n \in [2,5,6,7,8]$. The Gaussian windowed sum of Zernikes is then peak normalized as

\begin{equation}
\tilde{B}_k^{ng} = 0.2 \frac{B_i^{ng}}{\max\left(\left| B_i^{ng} \right| \right)}.
\end{equation}

We then sum the Gaussian and non-Gaussian components and integral normalize to form each
detector's beam,

\begin{equation}
B_k = \frac{B_k^g + \tilde{B}_k^{ng} }{\iint (B_k^g + \tilde{B}_k^{ng})}.
\end{equation}

\noindent The discretized beams are defined on a grid with spacing
$0.04\deg$ and size $2\deg \times 2\deg$.

\subsection{Scan Trajectory}\label{sec:traj}

We simulate TODs, including \lcdm\ + dust signal (``noiseless sims''),
\ttp\ leakage from beam mismatch, and instrument noise. We simulate these three
components separately because the subsequent mapmaking is a linear operation,
and so summing maps of the individual components is equivalent to simulating them
simultaneously in the time domain. (We have verified this numerically.)

We first generate a mock scan strategy for a single BICEP Array or CMB-Stage
IV-like small 
aperture telescope (SAT) \citep{BA1,BA2,CMBS4-ref}, which
defines the telescope boresight's pointing and azimuthal orientation as a
function of time. The scan strategy is simplified to keep the number of TOD
samples to a minimum, allowing us to try out many different deprojection
techniques using only a small amount of computing time. The main simplification
is that we simulate only a single ``scan'' at each elevation instead of
multiple back-and-forth scans, a simplification that should not impact
deprojection.  We tune the trajectory to yield a map with $f_{sky}$ roughly
appropriate for next-generation CMB surveys given the instantaneous
field-of-view of the focal plane (FP) defined below. We simulate
the boresight trajectory on a grid of R.A. and Dec.\ spanning $[-28.5\deg,
  +28.5\deg]$ and $[-57.5\deg, -32.5\deg]$, respectively. The step size in
Dec.\ is $0.25\deg$, chosen to be consistent with \bk. The step size in
(coordinate) R.A. is $0.35\deg$.

The scan trajectory is a single scan in azimuth at each Dec.\ followed by a step
in Dec.\ of $0.25\deg$, followed by a scan backward at the new Dec. In reality,
multiple scans would be made at each Dec.\ before stepping, allowing, for
instance, for azimuth-fixed template subtraction. We simulate the same scan
trajectory for 8 separate boresight orientations separated by $45\deg$,
i.e. $\{0\deg, 45\deg, 90\deg, ..., 315\deg\}$.

We then define orthogonally polarized detector-pair centroids relative to the
telescope boresight. We simulate a vastly reduced number of detectors, again in
order to be able to reduce computation time. To simulate an approximately
correct instantaneous field-of-view, we generate pair-centroids on a
$16\times16$ grid that is $28\deg\times28\deg$ with $1.86\deg$ spacing. We then
simulate each 10th pair so that we simulate only 26 detector pairs. The
detectors within a pair, which we label A and B, are defined to have orthogonal
polarization angles, with the A detector of the pair pointing vertically in FP
coordinates.

Reducing the detector count is valid because of the linearity of the later map
making and deprojection steps, so that a small number of low noise detectors is
qualitatively equivalent to a large number of noisy detectors. We just need to
sufficiently sample the FP so as to produce a smoothly apodized map.  A possible qualitative
difference when simulating fewer detectors is the net \ttp\ after coadding over
detectors. Beam mismatch that is truly random from detector-to-detector averages
down when coadding over detectors. Simulating a smaller number of detectors
underestimates this effect and therefore overestimates the net
contamination prior to deprojection. This should only set a higher bar than in
reality for the current study.

\begin{figure}[!t]
  \begin{center}
    \begin{tabular}{c}
      \includegraphics[width=1\linewidth]{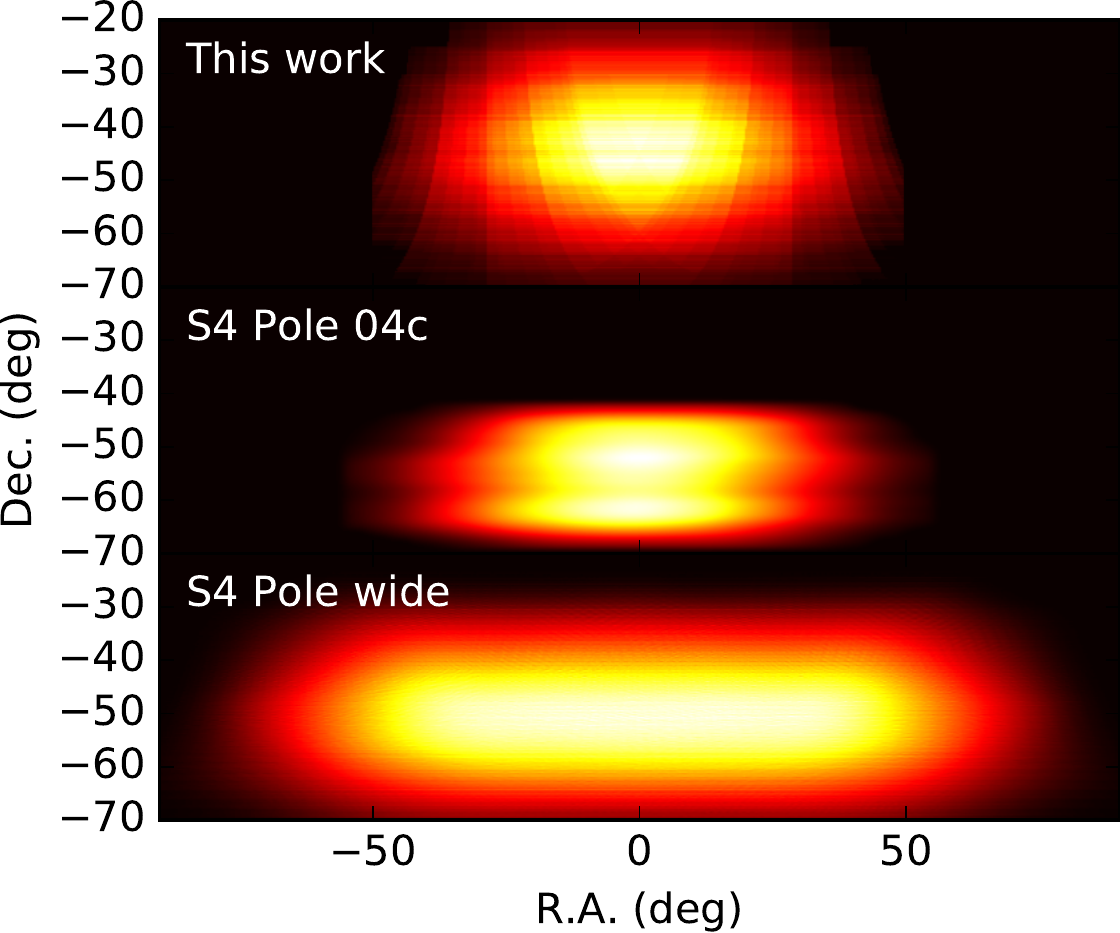}
    \end{tabular}
  \end{center}
  \caption[example]{ \label{fig:weight} Top: weight map for the scan trajectory
    and FP layout defined in this work. Middle: weight map for the deep Pole
    field (``experiment definition 04c'') from the CMB-S4 $r$-forecasting
    group. Bottom : weight map for the wide Pole field, centered in this plot on
    R.A. $= 0\deg$ to aid visual comparison, rather than R.A. $=30\deg$ as it
    could be in reality to avoid galactic foregrounds.}
\end{figure}

Each detector's centroid and polarization angle as projected onto the sky is
then computed as a function of time given the boresight pointing, boresight
rotation angle, and detector FP coordinate. Figure~\ref{fig:weight} shows the
polarization weight map generated by our mock scan strategy and FP layout, which
is equivalent to an $N_{hits}$ map because we simulate all detectors as having
identical noise that is constant in time.

\subsection{TOD Simulation}

As described in Section~\ref{sec:beams}, the mock beams are defined on a
regular Cartesian grid, $(x_j,y_j)$. We convert these to polar coordinates,
$(r_j,\phi_j)$, which we define as offsets in spherical coordinates 
relative to the centroid when the beam is projected onto the sky. Given the trajectory of the $k$th
detector's centroid defined in Section~\ref{sec:traj}, we compute the R.A. and
Dec.\ trajectory of each pixel $j$ in the discretized beam. We then sample the
simulated maps described in Section~\ref{sec:inputmaps} along these trajectories
using the \texttt{get\_interp\_val} routine of \texttt{Healpy}
to form time streams $T_{k,j}(t)$, $Q_{k,j}(t)$ and $U_{k,j}(t)$. Multiplying
each time stream by the value of the beam and summing yields the beam convolved
$T$, $Q$, and $U$ trajectories for detector $k$,

\begin{equation}\label{eq:Tk}
  T_k(t) = \sum_j B_k(r_j,\phi_j) T_{k,j}(t),
\end{equation}

\noindent and similarly for $Q_k(t)$ and $U_k(t)$. The observed TOD for the
$k$th detector is then computed as

\begin{equation}\label{eq:sk}
s_k(t) =T_k(t) + Q_k(t)\cos[2\chi_k(t)] + U_k(t)\sin[2\chi_k(t)]
\end{equation}

\noindent where $\chi_k(t)$ is the polarization angle of the detector as a
function of time. We then compute the pair-difference time stream of
the $i$th pair by differencing the A and B detector time streams within a pair,

\begin{equation}
d_i(t) = [s_{k=2i}(t) - s_{k=2i+1}(t)]/2 .
\end{equation}

We produce noiseless \lcdm\ as well as $E$-only simulations by sampling off the
maps described in Section~\ref{sec:inputmaps}. We also
produce ``leakage'' sims by setting $Q_k(t) = U_k(t) = 0$, so that the
only non-zero component of $d_i(t)$ must be due to \ttp\ leakage from beam
mismatch (i.e. $B_{2i} \ne B_{2i+1}$).

Lastly, we produce noise simulations by replacing the TOD with random Gaussian numbers
of mean zero and a width tuned to produce next-generation survey noise levels after coadding
over detectors.

\begin{figure}[!t]
  \begin{center}
    \begin{tabular}{c}
      \includegraphics[width=.7\linewidth]{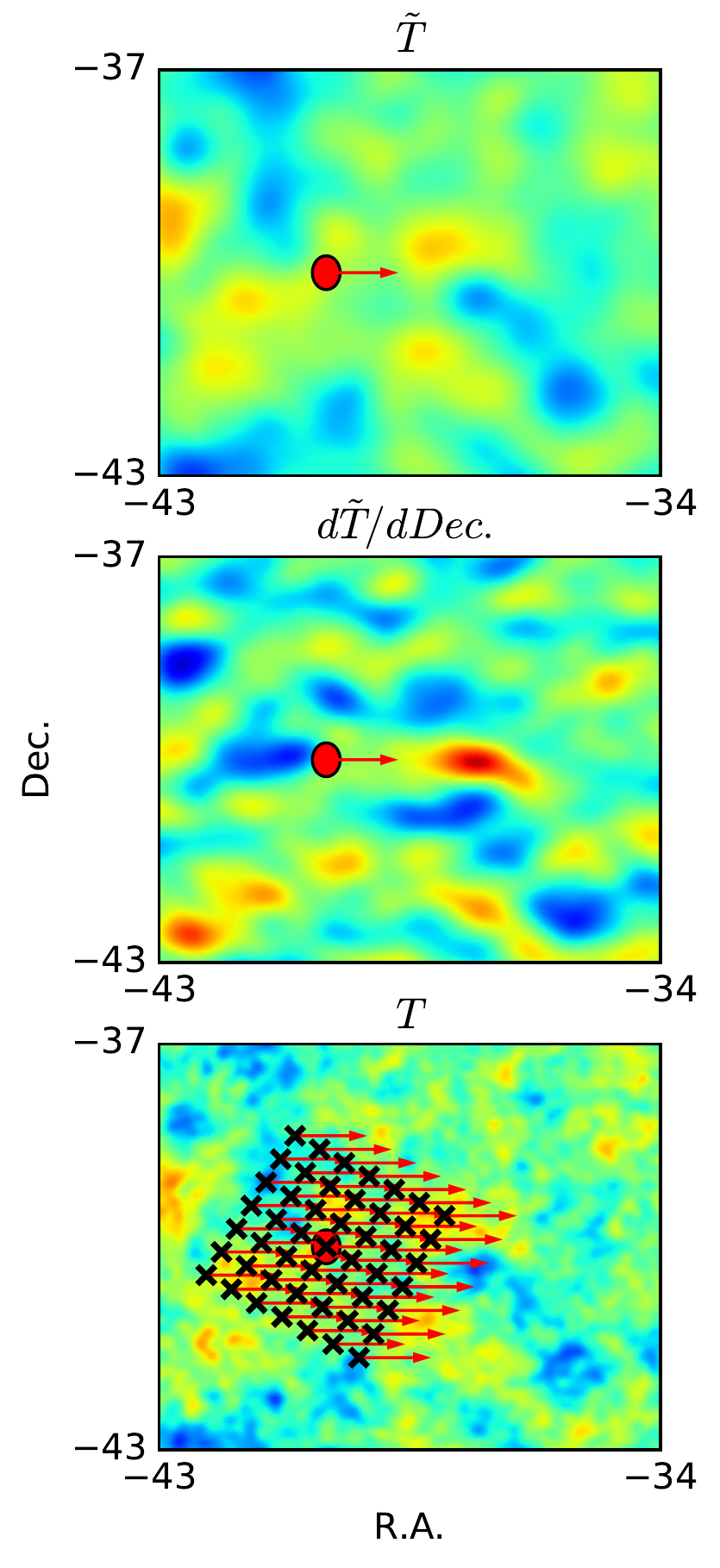}
    \end{tabular}
  \end{center}
  \caption[example]{ \label{fig:cartoon} Illustration of the construction of
    elliptical Gaussian leakage templates (top and middle) and arbitrary leakage
    templates (bottom). The top panel shows the nominal-beam-smoothed
    CMB temperature map, $\tilde{T}$, used to construct the differential gain
    template. The red circle
    illustrates a single detector-pair. The red arrow illustrates a short
    section of the trajectory of the
    detector-pair centroid during a typical fixed-elevation
    scan at the South Pole. The middle panel shows the smoothed first derivative
    map used to construct the differential pointing leakage template.
    The bottom panel shows the unsmoothed $T$ map and
    the grid defining the $N$ arbitrary leakage template trajectories. (The grid
    point spacing has been increased from $0.1\deg$ to $0.4\deg$ for visual
    clarity.)  The rotation of the grid about the beam centroid is for one of
    the 8 boresight angles.}
\end{figure}

\section{Mapmaking and Deprojection}
\label{sec:mapmaking}

\begin{figure*}[!ht]
  \begin{center}
    \begin{tabular}{c}
      \includegraphics[width=1\linewidth]{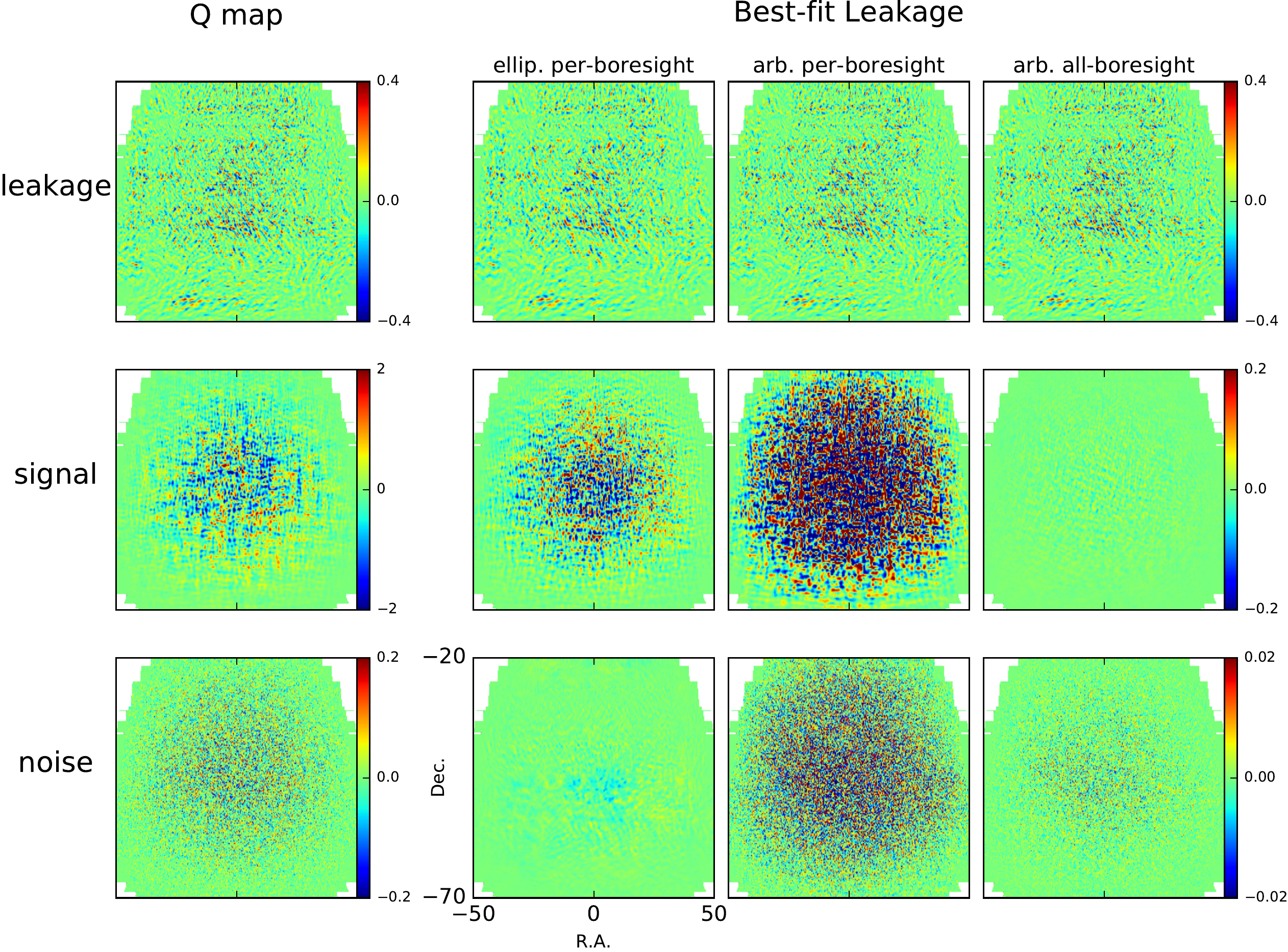}
    \end{tabular}
  \end{center}
  \caption[example]{ \label{fig:dp} Simulated $Q$ maps and best-fit leakage
    templates, broken down into simulation components and apodized by the weight
    map in the top panel of Figure~\ref{fig:weight}. (N.B., the aspect ratio of
    the maps in this plot is not faithful.)
    Rows from top to bottom
    show the three simulation components: \ttp\ leakage, \lcdm\ + dust signal,
    and instrumental noise. The leftmost column shows the
    simulated $Q$ map, coadded over all detectors and boresight angles, prior to
    deprojection. The remaining three columns show the best-fit leakage
    templates for different deprojection options. The ``ellip.\ per-boresight''
    column shows the template for elliptical Gaussian deprojection in which
    leakage templates are fit to data from each boresight angle separately. The
    ``arb.\ per-boresight'' column shows the arbitrary
    deprojection best-fit template, also fitting to each boresight angle separately. The
    ``arb.\ all-boresight'' column shows the arbitrary deprojection template
    when fitting all boresight angles simultaneously. The
    $Q$ maps on the left would be deprojected by subtracting one of the
    corresponding three maps on the right.}
\end{figure*}

We then bin the TODs into maps. We also construct time-ordered templates of
\ttp\ leakage and bin each of these into maps as well. We fit the these binned
leakage templates to the data and subtract them to deproject contamination from
beam shape mismatch. Overfitting will also filter some ``true'' signal. The
goal is to minimize this by keeping the complexity of the model to a minimum.

\subsection{Mapmaking}

We bin the simulated TODs into maps based on the procedure in~\cite{BKI}.
Like the \bk\ maps, the map pixel
boundaries are defined as a grid in R.A.\ and Dec.\ with size $0.25\deg$ in
Dec.\ and size in R.A.\ chosen to make the pixels square along the central row.

For each pair $i$, we bin into map pixels various products of: (1) the pair
difference TOD, $d_i(t)$; (2) the instantaneous weight, $w_i(t)$, which here we
set to 1 for all time samples for all pairs; and (3) $\sin2\chi_i$ and
$\cos2\chi_i$, where $\chi$ is the A detector's polarization angle projected
onto the sky.  We bin those products needed to later reconstruct $Q$ and
$U$. The binned quantity for each detector pair is referred to as a ``pair-map''
and is denoted

\begin{equation}\label{eq:pairmap}
m_{i}(\alpha_p, \delta_p) = \sum_l q_i(t_l),
\end{equation}

\noindent where the sum is over time samples for which the $i$th detector-pair's
centroid lies within the boundaries of pixel $p$ (which is centered on R.A. $\alpha_p$ and
Dec. $\delta_p$), and 

\begin{eqnarray}
q_i &  \subset & [w_i, w_id_i, w_id_i\cos2\chi_i, w_id_i\sin2\chi_i, \\
& &  \cos^2 2\chi_i, \sin^2 2\chi_i, \cos2\chi_i\sin2\chi_i].\nonumber
\end{eqnarray}

This set of $q_i$ are the minimum quantities needed to later solve for $Q$, and
$U$. They also have the property that they may be coadded over time, over
detectors, and over polarization angles. Technically, the quantity $w_id_i$ is
not needed to construct the $Q$ and $U$ maps, but we later fit systematics
templates to it. We bin data from different boresight angles separately so that
there is a separate pair-map for each angle.  The final step is to coadd over
boresight angles and solve for $Q$
and $U$, the mathematical details of which we do not give here.

\subsection{Leakage template construction}

We also generate templates of \ttp\ leakage.
In general, we model the \ttp\ leakage in $d_i(t)$ as 
the linear combination of $N$ ``leakage templates'', $l_{i,n}$, so that the
total contamination is

\begin{equation}\label{eq:lt}
  L_i(t) = \sum_{n=1}^N a_{i,n}l_{i,n}(t).
\end{equation}

Leakage templates are simultaneously fit to the data at either the time-ordered
level or the pair-map level to determine the $a_{i,n}$. BKIII fits leakage
templates directly to $d_i(t)$ on timescales that include contiguous
observations from a single boresight angle. This has the benefit of removing any
leakage that varies on timescales longer than this. There is good reason to
expect, however, that the beam is constant in time over a season, and indeed there is
no evidence for temporal or boresight angle beam dependence in \bk\ science or
calibration data. We therefore opt to first coadd the data into pair-maps
without first deprojecting, and fit the similarly coadded leakage
templates in map-space. We coadd each time-ordered leakage template by
binning it into map pixels like the data, 

\begin{equation}\label{eq:pairmap}
m_{i,n}^{T\rightarrow P}(\alpha_p, \delta_p) = \sum_l q_{i,n}(t_l),
\end{equation}

\noindent with

\begin{equation}
q_{i,n} \subset [w_il_{i,n}, w_il_{i,n}\cos2\chi_i, w_il_{i,n}\sin2\chi_i].
\end{equation}

For each detector pair, we fit the $N$ binned $w_il_{i,n}$ leakage templates to
the $w_id_i$ pair-map to find the $a_{i,n}$. We then multiply the
$w_il_{i,n}\sin2\chi_i$ and $w_il_{i,n}\cos2\chi_i$ leakage templates by the
best-fit $a_{i,n}$, sum over $n$, and store this single best-fit leakage
pair-map for that detector pair. Lastly, we coadd the summed
leakage templates over detector-pairs to construct the best-fit $Q$ and $U$
leakage templates, which we subtract from the maps.

Because we make a separate pair-map for each boresight angle, leakage does not
cancel in the maps prior to fitting. As noted, deprojecting in pairmap-space
will not perfectly remove mismatch that varies with time, as might be expected
from detector gain mismatch. We therefore envision that differential gain
deprojection will first be done at the time stream level, as is currently done by
\bk.

We alternately deproject two kinds of leakage templates: elliptical Gaussian
leakage templates like those described in BKIII, which model the beams as
elliptical Gaussians, and ``arbitrary'' leakage templates, which make no
assumptions about the beam shape.  BKIII simultaneously fits the ($N=6$)
$l_{i,n}$ corresponding to the modes of a mismatched elliptical Gaussian
directly to $d_i$. We form the same $l_{i,n}$ as in BKIII. Briefly summarized,
we construct them by first smoothing the (simulated) \planck\ $T$ map to the
nominal beam, a circular Gaussian of width FWHM$=0.5\deg$. We then sample
this map along with its first and second derivatives along the detector-pair's
trajectory. We then form the linear combinations of these derivative map
time streams that approximate the leakage from the modes of mismatched elliptical
Gaussians. There is one mode for amplitude mismatch (i.e. differential gain), two
for centroid (differential pointing), one for beam width, and two for
ellipticity. Although we bin these leakage templates into pair-maps prior to
fitting to data (unlike BKIII which fits in the time domain), we still fit data
from each boresight angle separately. Thus, while the total number of data
points being fit is greatly reduced relative to BKIII, the fractional degrees of
freedom being removed from the map is comparable. Another difference with the
implementation of deprojection by \bk\ is that we fit for the differential
ellipticity coefficients. The \bk\ results all set these coefficients to values
determined from external calibration data. The reason for this is that fitting
for these coefficients strongly filters the $TE$ spectrum and somewhat filters
the $EE$ spectrum, a phenomenon arising from the inherent correlation between
the $T$-map-derived leakage templates and the underlying \lcdm\ \emode\ signal.

In addition to the elliptical Gaussian templates, we also form deprojection
templates that make no assumption about beam shape. At each time sample in the
TOD, we form a gnomonic projection of the unsmoothed simulated Planck $T$ map
centered on the detector-pair centroid.  We define the projection grid with
respect to FP coordinates $\{x,y\}$ so that the projection rotates with the
projected orientation of the FP on the sky. The time series for a given grid
point is the leakage template, $l_{i,n}(t)$.  The projection size is $2.4\deg
\times 2.4\deg$ with a grid spacing of $0.1\deg$, so that the number of grid
points is $N=576$. When fitting the binned arbitrary leakage templates to the
data, we alternately fit data from each boresight angle separately to allow
direct comparison with differential ellipticity deprojection, and fit all
boresight angles simultaneously to reduce the filtering of the true signal.

Figure~\ref{fig:cartoon} illustrates the difference between the elliptical
Gaussian and arbitrary leakage templates. The top two panels illustrate
differential ellipticity template construction, showing a single detector-pair's
trajectory along the SAT-beam-smoothed $T$ map and one of its derivatives. (The
map is zoomed in for clarity, and the other derivative maps are omitted for
space.)  The pair's differential gain leakage template is constructed by
sampling the smoothed $T$ map along this trajectory. The leakage template for
the North-South component of differential pointing would be constructed by
sampling the first derivative map off the same trajectory. The bottom panel
illustrates arbitrary template construction, with multiple trajectories along
the single unsmoothed $T$ map. The grid shown in Figure~\ref{fig:cartoon} is the
same size as used in this work ($2.4\deg \times 2.4\deg$) but the spacing shown
is $0.4\deg$ instead of $0.1\deg$ for clarity. In principle, the $a_{i,n}$ for
the arbitrary templates can be chosen to construct the smoothed first derivative
map, as well as any other derivatives, or linear combinations thereof.

\section{Results}
\label{sec:results}

Lastly, we deproject the simulated \ttp\ leakage. We
show the results in maps and in their angular power
spectra.

\subsection{Results in map space}

\begin{figure}
  \begin{center}
    \begin{tabular}{c}
      \includegraphics[width=1\linewidth]{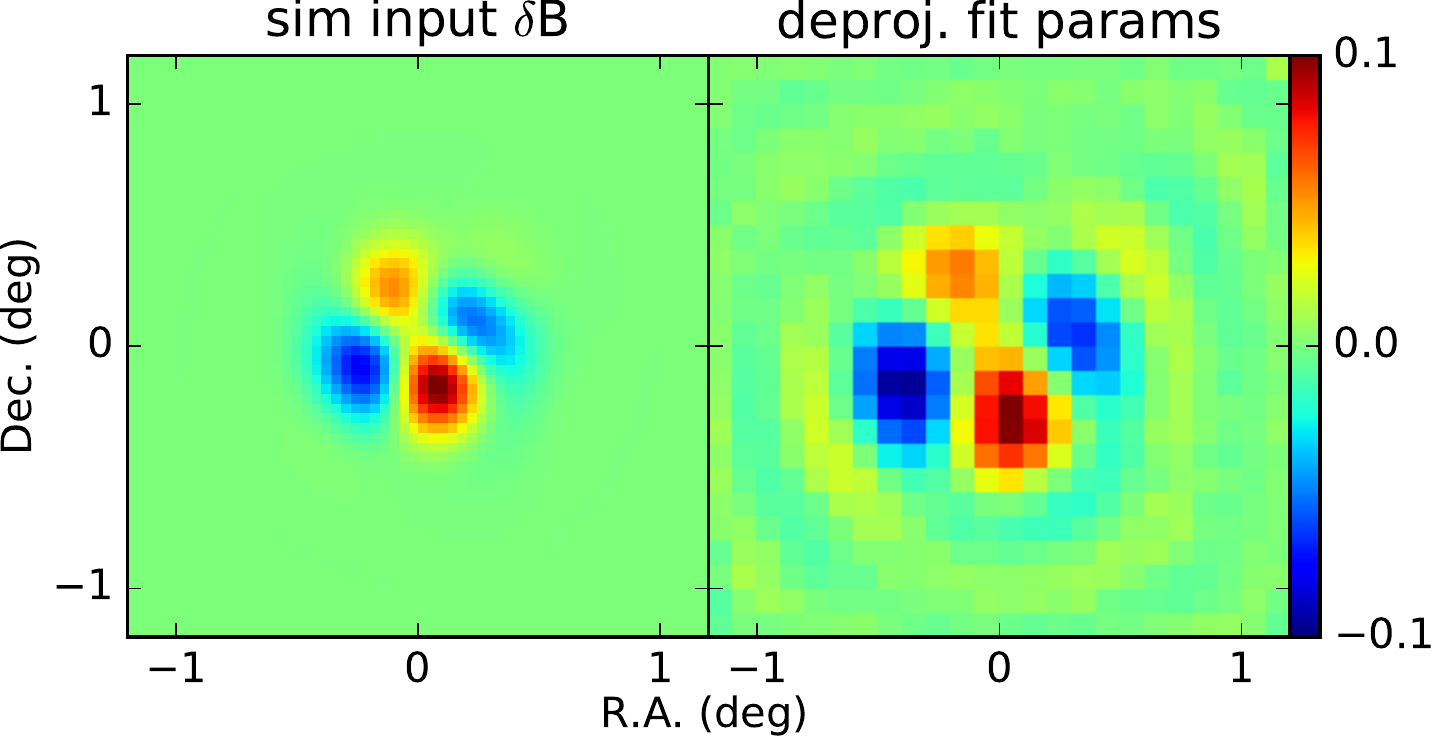} 
    \end{tabular}
  \end{center}
  \caption[example]{ \label{fig:beam} Left: the simulated difference beam for a single
    peak normalized detector pair. Right: 
    best-fit arbitrary deprojection coefficients for this detector pair in a
    single realization of the leakage-only simulations.}
\end{figure}

The $Q$ map for a single realization, broken down by simulation input component,
and the corresponding best-fit leakage templates under different deprojection
schemes are shown in Figure~\ref{fig:dp}. Each row is one of the three
components of the simulation: \ttp\ leakage, \lcdm\ + dust signal, and noise. The
left column shows the simulated maps ``as observed'' by our mock survey.  The
other columns show the best-fit leakage templates for different deprojection
options. Maps and leakage templates are coadded over all boresight angles and
detectors.

The top row shows the \ttp\ leakage component. Because the simulated beams
described in Section~\ref{sec:beams} were intentionally constructed to be close to
(but not quite) elliptical Gaussians, the best-fit templates are nearly
indistinguishable both from each other and from the actual leakage. As will be
shown later, however, the residuals are much lower for the arbitrary templates. 

The middle and bottom rows shows the signal and noise components,
respectively. Any non-zero values in these best-fit leakage templates are due to
overfitting and result in filtering of the true signal when subtracted from the
maps.  The elliptical Gaussian template, shown in the second column, establishes
a baseline level of overfitting that is \textit{de facto} acceptable for
Stage-III and Stage-IV experiments. This is because elliptical Gaussian
deprojection is used by \bk\ for all its $r$-constraints, and because the CMB-S4
inflation projections use the achieved \bk\ $BB$ noise levels as the starting
point for scaling to Stage-IV detector counts~\citep{CMBS4-ref}.
Any reduction of S/N caused by
preferentially filtering signal is already built into these projections. It
should be noted that elliptical Gaussian deprojection as actually implemented by
\bk\ (and therefore baselined by CMB-S4) is slightly different than what is
implemented here. \bk\ fits for only 4 of the 6 differential beam parameters;
the differential ellipticity coefficients are fixed to values measured from beam
maps. This slightly reduces the amount of overfitting in the \bk\ analysis is
therefore not perfectly comparable to our implementation of elliptical Gaussian
deprojection, which fits for all parameters. Nonetheless, \cite{barkats13}
demonstrates that filtering from differential gain deprojection is negligible
compared to azimuth-fixed template subtraction, which suppresses power by
$\sim50\%$ in the lowest \bk\ $EE$ bandpower. Since azimuth-fixed template
subtraction is implicitly baselined in the CMB-S4 projections, it should be a
safe assumption that demonstrating the absence of filtering in excess of
elliptical Gaussian deprojection is sufficient to demonstrate the suitability of
arbitrary deprojection for next-generation surveys.

Arbitrary deprojection on a per-boresight angle basis results in greater
overfitting relative to elliptical Gaussian deprojection, as shown by the larger
amplitude of the ``arb.\ per-boresight angle'' template relative to the
``ellip.\ per-boresight angle'' template in the middle and bottom rows.  This is
unsurprising given that the number of templates being fit is 576 vs. 6.

When fitting the arbitrary templates to all 8 boresight angles simultaneously,
however, as shown in the the ``arb.\ all-boresight angle'' column, the
overfitting of the signal component is \textit{less} than with ellipticity
deprojection. The overfitting of the noise component is somewhat higher at small
scales, though in fact is comparable at degree scales. This result is perhaps
surprising because, even though the number of map pixels being fit is $\approx
8\times$ greater relative to elliptical Gaussian deprojection, the
number of templates being fit is $\approx100\times$ greater. The naive
expectation would be that more degrees of freedom would be removed from the map.
We interpret this result as a consequence of correlations between the arbitrary
leakage templates. Generally treated
as a potential problem in regression, the reduction of effective degrees of
freedom in the leakage model of Equation~\ref{eq:lt} works in our favor. In
fact, since the correlation length of the CMB is $\sim1\deg$, then there are
$\sim6$ independent modes in our $2.4\deg \times 2.4\deg$ grid of leakage
templates, regardless of grid spacing.

\begin{figure*}[!ht]
  \begin{center}
    \begin{tabular}{c}
      \includegraphics[width=1\linewidth]{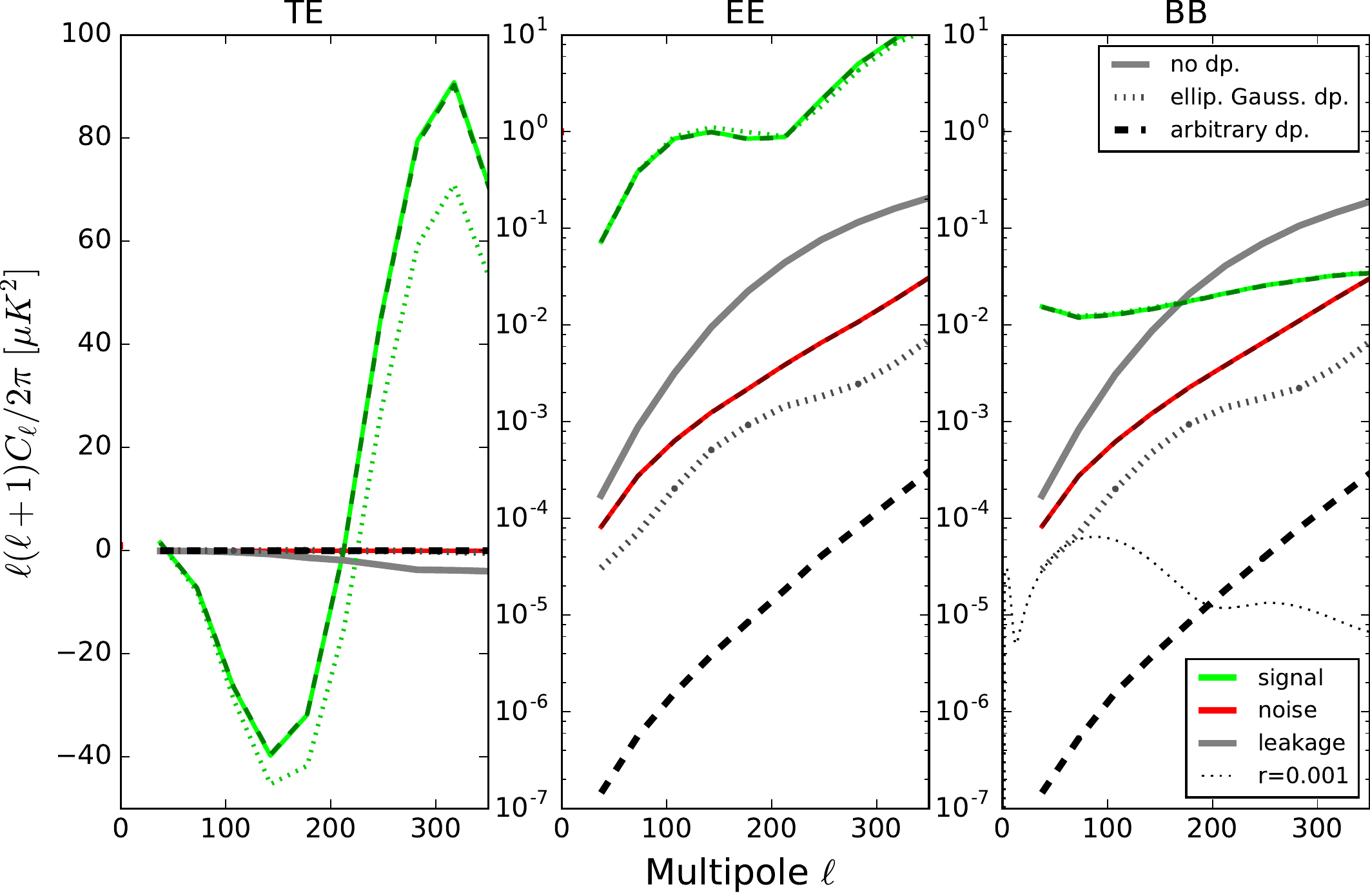}
    \end{tabular}
  \end{center}
  \caption[example]{ \label{fig:aps} Angular power spectra of the simulated
    maps before and after deprojection. The green, red, and gray lines show the
    signal, noise, and \ttp\ leakage components, respectively. The solid,
    dotted, and dashed lines show the spectra before deprojection, after
    elliptical Gaussian deprojection, and after all-boresight arbitrary
    deprojection, respectively.}
\end{figure*}

Figure~\ref{fig:beam} shows the simulated difference beam for a single
detector-pair along with the best-fit $a_n$ from the leakage-only
simulation. The fit coefficients are normalized to have the same peak value as
the input difference beam.  The two show good qualitative agreement, but there
is an apparent rescaling of the best-fit beam. We suspect that this is the
combined effect of \planck\ noise and correlations among templates.
Intuitively, there is a degeneracy between the angular separation of features in
the best-fit beam and their amplitude. For instance, differential pointing
manifests as a dipolar difference beam, which couples to the first derivative of
the $T$ map. To third order, a perturbation to the centroid offset within a
detector-pair can be approximated by either an increase in the dipole separation
or an increase in its amplitude.  While it is possible to construct a forward
simulation, mapmaking, and deprojection pipeline in which the known differential
beam is exactly recovered, we purposefully do not perform our forward simulation
in exactly same way as the template generation. We also include the effect of
Planck $T$ map noise in the template. We therefore do not necessarily expect
perfect agreement between the recovered fit coefficients and the input
beam. 

In principle, multicollinearity among regressors can lead to biased predictions
of the leakage if the coefficients derived from a regression on one data set are
used to generate predictions of leakage in a disjoint data set. Deprojection
both fits and predicts leakage using the same data so this is not an issue for
the current work. Furthermore, to the extent it could be an issue, our
simulations include it. What would perhaps be a problem is if coefficients
determined from fits to data whose dominant signal has a different scale length
than the CMB (like, say, galactic foregrounds)
were used to predict leakage in a CMB-dominated field. In this case,
our method might yield a biased prediction and would need to be tested on simulations.

\subsection{Results in Angular Power Spectra}

Lastly, we subtract the best-fit leakage templates from the maps and compute
the resulting angular power spectra. We again do so for the three simulated components
separately. Angular power spectra are calculated
as in \cite{BKI}. Briefly, we multiply the maps by the weight map, zero pad,
compute the 2D power spectrum with an FFT, and bin into 1D bandpowers. We account
for \etb\ mixing by computing the \bmode\ 2D Fourier plane of the $E$-only sims
and subtracting these modes directly in the Fourier plane of the \lcdm\ + dust
sims. This, of course, is not an option in reality since one can not know the
pure-E component ahead of time.  ``Pure-B'' estimators that do not leak
\etb\ for filtered maps are used to avoid this~\citep{BKVII}, but we have not implemented
one, and so use this simulation-friendly workaround.
Lastly, we correct the bandpowers for beam suppression by a circular
Gaussian using the appropriate analytic beam window function.

Figure~\ref{fig:aps} shows the angular power spectra of the coadded
maps prior to deprojection, after ellipticity deprojection, and after
all-boresight-angle arbitrary deprojection. 
We unsurprisingly find that arbitrary beam deprojection removes almost all of
the leakage while elliptical Gaussian deprojection leaves a significant
undeprojected residual. The size of this residual is, of course, dependent on the
component of the simulated beams that is not modeled by an elliptical
Gaussian, something we are free to choose in our simulations. We chose
the beams so that the residual with elliptical Gaussian deprojection would be small but important for
next-generation CMB surveys, which report projected sensitivities in
the neighborhood of $\sigma_r = 5\times10^{-4}$. The undeprojected residual with
arbitrary deprojection is of the order $r\approx 1\times10^{-5}$, more than
sufficient for the future.

The filtering of all true signal appears negligible. Happily, arbitrary
deprojection does not appreciably filter $EE$ and $TE$ like ellipticity
deprojection does. We therefore conclude that our method is a success.

\section{Conclusions}
\label{sec:conclusion}

We conclude that arbitrary deprojection, performed over all boresight angles
simultaneously, solves the problem of \ttp\ leakage from main beam mismatch for
next-generation CMB surveys. Our templates extend out to a distance of
$1.2\deg$ from the beam centroids, so leakage arising from near and far sidelobes
will not be cleaned by this procedure. Possibly, the templates can be extended to
larger distances to encompass near sidelobes. Extending the templates to
encompass far sidelobes will probably result in too much filtering of signal. It
may be possible to construct a second set of coarsely gridded templates from
smoothed $T$ maps to deproject sidelobe leakage under the assumption that
sidelobes have smoother features than the main beam. Nonetheless, beam maps of
sidelobes will almost certainly remain a crucial component of next-generation CMB
surveys. Arbitrary deprojection of main beam leakage can allow more
effort to be placed here.

\bibliography{ms}

\begin{thebibliography}{15}%
\makeatletter
\providecommand \@ifxundefined [1]{%
 \@ifx{#1\undefined}
}%
\providecommand \@ifnum [1]{%
 \ifnum #1\expandafter \@firstoftwo
 \else \expandafter \@secondoftwo
 \fi
}%
\providecommand \@ifx [1]{%
 \ifx #1\expandafter \@firstoftwo
 \else \expandafter \@secondoftwo
 \fi
}%
\providecommand \natexlab [1]{#1}%
\providecommand \enquote  [1]{``#1''}%
\providecommand \bibnamefont  [1]{#1}%
\providecommand \bibfnamefont [1]{#1}%
\providecommand \citenamefont [1]{#1}%
\providecommand \href@noop [0]{\@secondoftwo}%
\providecommand \href [0]{\begingroup \@sanitize@url \@href}%
\providecommand \@href[1]{\@@startlink{#1}\@@href}%
\providecommand \@@href[1]{\endgroup#1\@@endlink}%
\providecommand \@sanitize@url [0]{\catcode `\\12\catcode `\$12\catcode
  `\&12\catcode `\#12\catcode `\^12\catcode `\_12\catcode `\%12\relax}%
\providecommand \@@startlink[1]{}%
\providecommand \@@endlink[0]{}%
\providecommand \url  [0]{\begingroup\@sanitize@url \@url }%
\providecommand \@url [1]{\endgroup\@href {#1}{\urlprefix }}%
\providecommand \urlprefix  [0]{URL }%
\providecommand \Eprint [0]{\href }%
\providecommand \doibase [0]{http://dx.doi.org/}%
\providecommand \selectlanguage [0]{\@gobble}%
\providecommand \bibinfo  [0]{\@secondoftwo}%
\providecommand \bibfield  [0]{\@secondoftwo}%
\providecommand \translation [1]{[#1]}%
\providecommand \BibitemOpen [0]{}%
\providecommand \bibitemStop [0]{}%
\providecommand \bibitemNoStop [0]{.\EOS\space}%
\providecommand \EOS [0]{\spacefactor3000\relax}%
\providecommand \BibitemShut  [1]{\csname bibitem#1\endcsname}%
\let\auto@bib@innerbib\@empty
\bibitem [{\citenamefont {{\biceptwo\ Collaboration III}}(2015)}]{BKIII}%
  \BibitemOpen
  \bibfield  {author} {\bibinfo {author} {\bibnamefont {{\biceptwo\
  Collaboration III}}},\ }\href {\doibase 10.1088/0004-637X/814/2/110}
  {\bibfield  {journal} {\bibinfo  {journal} {\apj}\ }\textbf {\bibinfo
  {volume} {814}},\ \bibinfo {eid} {110} (\bibinfo {year} {2015})},\ \Eprint
  {http://arxiv.org/abs/1502.00608} {arXiv:1502.00608 [astro-ph.IM]}
  \BibitemShut {NoStop}%
\bibitem [{\citenamefont {{\biceptwo\ and \keckarray\ Collaborations
  IV}}(2015)}]{BKIV}%
  \BibitemOpen
  \bibfield  {author} {\bibinfo {author} {\bibnamefont {{\biceptwo\ and
  \keckarray\ Collaborations IV}}},\ }\href {\doibase
  10.1088/0004-637X/806/2/206} {\bibfield  {journal} {\bibinfo  {journal}
  {\apj}\ }\textbf {\bibinfo {volume} {806}},\ \bibinfo {eid} {206} (\bibinfo
  {year} {2015})},\ \Eprint {http://arxiv.org/abs/1502.00596} {arXiv:1502.00596
  [astro-ph.IM]} \BibitemShut {NoStop}%
\bibitem [{\citenamefont {{\biceptwo\ and \keckarray\ Collaborations
  XI}}(2019)}]{BKXI}%
  \BibitemOpen
  \bibfield  {author} {\bibinfo {author} {\bibnamefont {{\biceptwo\ and
  \keckarray\ Collaborations XI}}},\ }\href {\doibase 10.3847/1538-4357/ab391d}
  {\bibfield  {journal} {\bibinfo  {journal} {\apj}\ }\textbf {\bibinfo
  {volume} {884}},\ \bibinfo {eid} {114} (\bibinfo {year} {2019})},\ \Eprint
  {http://arxiv.org/abs/1904.01640} {arXiv:1904.01640 [astro-ph.IM]}
  \BibitemShut {NoStop}%
\bibitem [{\citenamefont {{Planck Collaboration}}\ \emph
  {et~al.}(2014)\citenamefont {{Planck Collaboration}}, \citenamefont {{Ade}},
  \citenamefont {{Aghanim}}, \citenamefont {{Armitage-Caplan}}, \citenamefont
  {{Arnaud}}, \citenamefont {{Ashdown}}, \citenamefont {{Atrio-Barandela}},
  \citenamefont {{Aumont}}, \citenamefont {{Baccigalupi}}, \citenamefont
  {{Banday}},\ and\ \citenamefont {et~al.}}]{P2013XVI}%
  \BibitemOpen
  \bibfield  {author} {\bibinfo {author} {\bibnamefont {{Planck
  Collaboration}}}, \bibinfo {author} {\bibfnamefont {P.~A.~R.}\ \bibnamefont
  {{Ade}}}, \bibinfo {author} {\bibfnamefont {N.}~\bibnamefont {{Aghanim}}},
  \bibinfo {author} {\bibfnamefont {C.}~\bibnamefont {{Armitage-Caplan}}},
  \bibinfo {author} {\bibfnamefont {M.}~\bibnamefont {{Arnaud}}}, \bibinfo
  {author} {\bibfnamefont {M.}~\bibnamefont {{Ashdown}}}, \bibinfo {author}
  {\bibfnamefont {F.}~\bibnamefont {{Atrio-Barandela}}}, \bibinfo {author}
  {\bibfnamefont {J.}~\bibnamefont {{Aumont}}}, \bibinfo {author}
  {\bibfnamefont {C.}~\bibnamefont {{Baccigalupi}}}, \bibinfo {author}
  {\bibfnamefont {A.~J.}\ \bibnamefont {{Banday}}}, \ and\ \bibinfo {author}
  {\bibnamefont {et~al.}},\ }\href {\doibase 10.1051/0004-6361/201321591}
  {\bibfield  {journal} {\bibinfo  {journal} {\aap}\ }\textbf {\bibinfo
  {volume} {571}},\ \bibinfo {eid} {A16} (\bibinfo {year} {2014})},\ \Eprint
  {http://arxiv.org/abs/1303.5076} {arXiv:1303.5076} \BibitemShut {NoStop}%
\bibitem [{\citenamefont {{Zaldarriaga}}\ and\ \citenamefont
  {{Seljak}}(1998)}]{zaldarriaga98}%
  \BibitemOpen
  \bibfield  {author} {\bibinfo {author} {\bibfnamefont {M.}~\bibnamefont
  {{Zaldarriaga}}}\ and\ \bibinfo {author} {\bibfnamefont {U.}~\bibnamefont
  {{Seljak}}},\ }\href {\doibase 10.1103/PhysRevD.58.023003} {\bibfield
  {journal} {\bibinfo  {journal} {\prd}\ }\textbf {\bibinfo {volume} {58}},\
  \bibinfo {eid} {023003} (\bibinfo {year} {1998})},\ \Eprint
  {http://arxiv.org/abs/astro-ph/9803150} {astro-ph/9803150} \BibitemShut
  {NoStop}%
\bibitem [{\citenamefont {{Planck Collaboration}}\ \emph
  {et~al.}(2016{\natexlab{a}})\citenamefont {{Planck Collaboration}},
  \citenamefont {{Adam}}, \citenamefont {{Ade}}, \citenamefont {{Aghanim}},
  \citenamefont {{Arnaud}}, \citenamefont {{Aumont}}, \citenamefont
  {{Baccigalupi}}, \citenamefont {{Banday}}, \citenamefont {{Barreiro}},
  \citenamefont {{Bartlett}},\ and\ \citenamefont {et~al.}}]{PIPXXX}%
  \BibitemOpen
  \bibfield  {author} {\bibinfo {author} {\bibnamefont {{Planck
  Collaboration}}}, \bibinfo {author} {\bibfnamefont {R.}~\bibnamefont
  {{Adam}}}, \bibinfo {author} {\bibfnamefont {P.~A.~R.}\ \bibnamefont
  {{Ade}}}, \bibinfo {author} {\bibfnamefont {N.}~\bibnamefont {{Aghanim}}},
  \bibinfo {author} {\bibfnamefont {M.}~\bibnamefont {{Arnaud}}}, \bibinfo
  {author} {\bibfnamefont {J.}~\bibnamefont {{Aumont}}}, \bibinfo {author}
  {\bibfnamefont {C.}~\bibnamefont {{Baccigalupi}}}, \bibinfo {author}
  {\bibfnamefont {A.~J.}\ \bibnamefont {{Banday}}}, \bibinfo {author}
  {\bibfnamefont {R.~B.}\ \bibnamefont {{Barreiro}}}, \bibinfo {author}
  {\bibfnamefont {J.~G.}\ \bibnamefont {{Bartlett}}}, \ and\ \bibinfo {author}
  {\bibnamefont {et~al.}},\ }\href {\doibase 10.1051/0004-6361/201425034}
  {\bibfield  {journal} {\bibinfo  {journal} {\aap}\ }\textbf {\bibinfo
  {volume} {586}},\ \bibinfo {eid} {A133} (\bibinfo {year}
  {2016}{\natexlab{a}})},\ \Eprint {http://arxiv.org/abs/1409.5738}
  {arXiv:1409.5738} \BibitemShut {NoStop}%
\bibitem [{\citenamefont {{Sheehy}}\ and\ \citenamefont
  {{Slosar}}(2018)}]{Sheehy18}%
  \BibitemOpen
  \bibfield  {author} {\bibinfo {author} {\bibfnamefont {C.}~\bibnamefont
  {{Sheehy}}}\ and\ \bibinfo {author} {\bibfnamefont {A.}~\bibnamefont
  {{Slosar}}},\ }\href {\doibase 10.1103/PhysRevD.97.043522} {\bibfield
  {journal} {\bibinfo  {journal} {\prd}\ }\textbf {\bibinfo {volume} {97}},\
  \bibinfo {eid} {043522} (\bibinfo {year} {2018})},\ \Eprint
  {http://arxiv.org/abs/1709.09729} {arXiv:1709.09729} \BibitemShut {NoStop}%
\bibitem [{\citenamefont {{\biceptwo\ and \keckarray\ Collaborations
  X}}(2018)}]{BKX}%
  \BibitemOpen
  \bibfield  {author} {\bibinfo {author} {\bibnamefont {{\biceptwo\ and
  \keckarray\ Collaborations X}}},\ }\href {\doibase
  10.1103/PhysRevLett.121.221301} {\bibfield  {journal} {\bibinfo  {journal}
  {Physical Review Letters}\ }\textbf {\bibinfo {volume} {121}},\ \bibinfo
  {eid} {221301} (\bibinfo {year} {2018})},\ \Eprint
  {http://arxiv.org/abs/1810.05216} {arXiv:1810.05216} \BibitemShut {NoStop}%
\bibitem [{\citenamefont {{Planck Collaboration}}\ \emph
  {et~al.}(2016{\natexlab{b}})\citenamefont {{Planck Collaboration}},
  \citenamefont {{Ade}}, \citenamefont {{Aghanim}}, \citenamefont {{Arnaud}},
  \citenamefont {{Ashdown}}, \citenamefont {{Aumont}}, \citenamefont
  {{Baccigalupi}}, \citenamefont {{Banday}}, \citenamefont {{Barreiro}},
  \citenamefont {{Bartlett}},\ and\ \citenamefont {et~al.}}]{P2015XII}%
  \BibitemOpen
  \bibfield  {author} {\bibinfo {author} {\bibnamefont {{Planck
  Collaboration}}}, \bibinfo {author} {\bibfnamefont {P.~A.~R.}\ \bibnamefont
  {{Ade}}}, \bibinfo {author} {\bibfnamefont {N.}~\bibnamefont {{Aghanim}}},
  \bibinfo {author} {\bibfnamefont {M.}~\bibnamefont {{Arnaud}}}, \bibinfo
  {author} {\bibfnamefont {M.}~\bibnamefont {{Ashdown}}}, \bibinfo {author}
  {\bibfnamefont {J.}~\bibnamefont {{Aumont}}}, \bibinfo {author}
  {\bibfnamefont {C.}~\bibnamefont {{Baccigalupi}}}, \bibinfo {author}
  {\bibfnamefont {A.~J.}\ \bibnamefont {{Banday}}}, \bibinfo {author}
  {\bibfnamefont {R.~B.}\ \bibnamefont {{Barreiro}}}, \bibinfo {author}
  {\bibfnamefont {J.~G.}\ \bibnamefont {{Bartlett}}}, \ and\ \bibinfo {author}
  {\bibnamefont {et~al.}},\ }\href {\doibase 10.1051/0004-6361/201527103}
  {\bibfield  {journal} {\bibinfo  {journal} {\aap}\ }\textbf {\bibinfo
  {volume} {594}},\ \bibinfo {eid} {A12} (\bibinfo {year}
  {2016}{\natexlab{b}})},\ \Eprint {http://arxiv.org/abs/1509.06348}
  {arXiv:1509.06348} \BibitemShut {NoStop}%
\bibitem [{\citenamefont {{Hui}}\ and\ \citenamefont {{BICEP Array
  Collaboration}}(2018)}]{BA1}%
  \BibitemOpen
  \bibfield  {author} {\bibinfo {author} {\bibfnamefont {H.}~\bibnamefont
  {{Hui}}}\ and\ \bibinfo {author} {\bibnamefont {{BICEP Array
  Collaboration}}},\ }in\ \href {\doibase 10.1117/12.2311725} {\emph {\bibinfo
  {booktitle} {\procspie}}},\ \bibinfo {series} {Society of Photo-Optical
  Instrumentation Engineers (SPIE) Conference Series}, Vol.\ \bibinfo {volume}
  {10708}\ (\bibinfo {year} {2018})\ p.\ \bibinfo {pages} {1070807},\ \Eprint
  {http://arxiv.org/abs/1808.00568} {arXiv:1808.00568 [astro-ph.IM]}
  \BibitemShut {NoStop}%
\bibitem [{\citenamefont {{Crumrine}}\ and\ \citenamefont {{BICEP Array
  Collaboration}}(2018)}]{BA2}%
  \BibitemOpen
  \bibfield  {author} {\bibinfo {author} {\bibfnamefont {M.}~\bibnamefont
  {{Crumrine}}}\ and\ \bibinfo {author} {\bibnamefont {{BICEP Array
  Collaboration}}},\ }in\ \href {\doibase 10.1117/12.2312829} {\emph {\bibinfo
  {booktitle} {\procspie}}},\ \bibinfo {series} {Society of Photo-Optical
  Instrumentation Engineers (SPIE) Conference Series}, Vol.\ \bibinfo {volume}
  {10708}\ (\bibinfo {year} {2018})\ p.\ \bibinfo {pages} {107082D},\ \Eprint
  {http://arxiv.org/abs/1808.00569} {arXiv:1808.00569 [astro-ph.IM]}
  \BibitemShut {NoStop}%
\bibitem [{\citenamefont {{CMB-S4 Collaboration}}(2019)}]{CMBS4-ref}%
  \BibitemOpen
  \bibfield  {author} {\bibinfo {author} {\bibnamefont {{CMB-S4
  Collaboration}}},\ }\href@noop {} {\bibfield  {journal} {\bibinfo  {journal}
  {arXiv e-prints}\ ,\ \bibinfo {eid} {arXiv:1907.04473}} (\bibinfo {year}
  {2019})},\ \Eprint {http://arxiv.org/abs/1907.04473} {arXiv:1907.04473
  [astro-ph.IM]} \BibitemShut {NoStop}%
\bibitem [{\citenamefont {{\bicep2 Collaboration I}}(2014)}]{BKI}%
  \BibitemOpen
  \bibfield  {author} {\bibinfo {author} {\bibnamefont {{\bicep2 Collaboration
  I}}},\ }\href {\doibase 10.1103/PhysRevLett.112.241101} {\bibfield  {journal}
  {\bibinfo  {journal} {Physical Review Letters}\ }\textbf {\bibinfo {volume}
  {112}},\ \bibinfo {eid} {241101} (\bibinfo {year} {2014})},\ \Eprint
  {http://arxiv.org/abs/1403.3985} {arXiv:1403.3985} \BibitemShut {NoStop}%
\bibitem [{\citenamefont {{Barkats}}\ \emph {et~al.}(2014)\citenamefont
  {{Barkats}}, \citenamefont {{Aikin}}, \citenamefont {{Bischoff}},
  \citenamefont {{Buder}}, \citenamefont {{Kaufman}}, \citenamefont
  {{Keating}}, \citenamefont {{Kovac}}, \citenamefont {{Su}}, \citenamefont
  {{Ade}}, \citenamefont {{Battle}}, \citenamefont {{Bierman}}, \citenamefont
  {{Bock}}, \citenamefont {{Chiang}}, \citenamefont {{Dowell}}, \citenamefont
  {{Duband}}, \citenamefont {{Filippini}}, \citenamefont {{Hivon}},
  \citenamefont {{Holzapfel}}, \citenamefont {{Hristov}}, \citenamefont
  {{Jones}}, \citenamefont {{Kuo}}, \citenamefont {{Leitch}}, \citenamefont
  {{Mason}}, \citenamefont {{Matsumura}}, \citenamefont {{Nguyen}},
  \citenamefont {{Ponthieu}}, \citenamefont {{Pryke}}, \citenamefont
  {{Richter}}, \citenamefont {{Rocha}}, \citenamefont {{Sheehy}}, \citenamefont
  {{Kernasovskiy}}, \citenamefont {{Takahashi}}, \citenamefont {{Tolan}},\ and\
  \citenamefont {{Yoon}}}]{barkats13}%
  \BibitemOpen
  \bibfield  {author} {\bibinfo {author} {\bibfnamefont {D.}~\bibnamefont
  {{Barkats}}}, \bibinfo {author} {\bibfnamefont {R.}~\bibnamefont {{Aikin}}},
  \bibinfo {author} {\bibfnamefont {C.}~\bibnamefont {{Bischoff}}}, \bibinfo
  {author} {\bibfnamefont {I.}~\bibnamefont {{Buder}}}, \bibinfo {author}
  {\bibfnamefont {J.~P.}\ \bibnamefont {{Kaufman}}}, \bibinfo {author}
  {\bibfnamefont {B.~G.}\ \bibnamefont {{Keating}}}, \bibinfo {author}
  {\bibfnamefont {J.~M.}\ \bibnamefont {{Kovac}}}, \bibinfo {author}
  {\bibfnamefont {M.}~\bibnamefont {{Su}}}, \bibinfo {author} {\bibfnamefont
  {P.~A.~R.}\ \bibnamefont {{Ade}}}, \bibinfo {author} {\bibfnamefont {J.~O.}\
  \bibnamefont {{Battle}}}, \bibinfo {author} {\bibfnamefont {E.~M.}\
  \bibnamefont {{Bierman}}}, \bibinfo {author} {\bibfnamefont {J.~J.}\
  \bibnamefont {{Bock}}}, \bibinfo {author} {\bibfnamefont {H.~C.}\
  \bibnamefont {{Chiang}}}, \bibinfo {author} {\bibfnamefont {C.~D.}\
  \bibnamefont {{Dowell}}}, \bibinfo {author} {\bibfnamefont {L.}~\bibnamefont
  {{Duband}}}, \bibinfo {author} {\bibfnamefont {J.}~\bibnamefont
  {{Filippini}}}, \bibinfo {author} {\bibfnamefont {E.~F.}\ \bibnamefont
  {{Hivon}}}, \bibinfo {author} {\bibfnamefont {W.~L.}\ \bibnamefont
  {{Holzapfel}}}, \bibinfo {author} {\bibfnamefont {V.~V.}\ \bibnamefont
  {{Hristov}}}, \bibinfo {author} {\bibfnamefont {W.~C.}\ \bibnamefont
  {{Jones}}}, \bibinfo {author} {\bibfnamefont {C.~L.}\ \bibnamefont {{Kuo}}},
  \bibinfo {author} {\bibfnamefont {E.~M.}\ \bibnamefont {{Leitch}}}, \bibinfo
  {author} {\bibfnamefont {P.~V.}\ \bibnamefont {{Mason}}}, \bibinfo {author}
  {\bibfnamefont {T.}~\bibnamefont {{Matsumura}}}, \bibinfo {author}
  {\bibfnamefont {H.~T.}\ \bibnamefont {{Nguyen}}}, \bibinfo {author}
  {\bibfnamefont {N.}~\bibnamefont {{Ponthieu}}}, \bibinfo {author}
  {\bibfnamefont {C.}~\bibnamefont {{Pryke}}}, \bibinfo {author} {\bibfnamefont
  {S.}~\bibnamefont {{Richter}}}, \bibinfo {author} {\bibfnamefont
  {G.}~\bibnamefont {{Rocha}}}, \bibinfo {author} {\bibfnamefont
  {C.}~\bibnamefont {{Sheehy}}}, \bibinfo {author} {\bibfnamefont {S.~S.}\
  \bibnamefont {{Kernasovskiy}}}, \bibinfo {author} {\bibfnamefont {Y.~D.}\
  \bibnamefont {{Takahashi}}}, \bibinfo {author} {\bibfnamefont {J.~E.}\
  \bibnamefont {{Tolan}}}, \ and\ \bibinfo {author} {\bibfnamefont {K.~W.}\
  \bibnamefont {{Yoon}}},\ }\href {\doibase 10.1088/0004-637X/783/2/67}
  {\bibfield  {journal} {\bibinfo  {journal} {\apj}\ }\textbf {\bibinfo
  {volume} {783}},\ \bibinfo {eid} {67} (\bibinfo {year} {2014})},\ \Eprint
  {http://arxiv.org/abs/1310.1422} {arXiv:1310.1422} \BibitemShut {NoStop}%
\bibitem [{\citenamefont {{\biceptwo\ and \keckarray\ Collaborations
  VII}}(2016)}]{BKVII}%
  \BibitemOpen
  \bibfield  {author} {\bibinfo {author} {\bibnamefont {{\biceptwo\ and
  \keckarray\ Collaborations VII}}},\ }\href {\doibase
  10.3847/0004-637X/825/1/66} {\bibfield  {journal} {\bibinfo  {journal}
  {\apj}\ }\textbf {\bibinfo {volume} {825}},\ \bibinfo {eid} {66} (\bibinfo
  {year} {2016})},\ \Eprint {http://arxiv.org/abs/1603.05976} {arXiv:1603.05976
  [astro-ph.IM]} \BibitemShut {NoStop}%
\end{thebibliography}%

\end{document}